\def\tr{{\rm tr}\,}
\def\Tr{{\rm Tr}\,}
\def\sgn{{\rm sgn\,}}
\def\b{\bibitem}
\def\wt{\widetilde}
\def\be{\begin{equation}}
\def\ee{\end{equation}}
\def\bea{\begin{eqnarray}}
\def\eea{\end{eqnarray}}
\def\bml{\begin{mathletters}}
\def\eml{\end{mathletters}}
\def\m{\hskip -3pt}
\documentstyle[aps,prb,eqsecnum,psfig,floats]{revtex}
\begin{document}
\def\SNG{{\em Physical Review Style and Notation Guide}}
\def\LUG {{\em \LaTeX{} User's Guide \& Reference Manual}}
\def\btt#1{{\tt$\backslash$\string#1}}%
\def\REVTeX{REV\TeX}
\def\AmS{{\protect\the\textfont2
        A\kern-.1667em\lower.5ex\hbox{M}\kern-.125emS}}
\def\AmSLaTeX{\AmS-\LaTeX}
\def\BibTeX{\rm B{\sc ib}\TeX}
\twocolumn[\hsize\textwidth\columnwidth\hsize\csname@twocolumnfalse%
\endcsname
 
\title{Absence of electron dephasing at zero temperature\\
}
\author{T.R.Kirkpatrick}
\address{Institute for Physical Science and Technology, and Department of 
         Physics\\
         University of Maryland,\\ 
         College Park, MD 20742}
\author{D.Belitz}
\address{Department of Physics and Materials Science Institute\\
         University of Oregon,\\
         Eugene, OR 97403}
\date{\today}
\maketitle

\begin{abstract}
Dephasing of electrons due to the electron-electron interaction has
recently been the subject of a controversial debate, with different
calculations yielding mutually incompatible results. In this paper we 
prove, by means of Ward identities, that in a system with time
reversal invariance, neither a Coulomb interaction 
nor a short-ranged model interaction can lead to phase breaking at 
zero temperature in spatial dimensions $d>2$.

\end{abstract}
\pacs{PACS numbers:  }
]

\section{Introduction}
\label{sec:I}

Many interesting phenomena in electronic systems at low temperatures depend
on the electrons' phase coherence. Among these are the `weak-localization
effects',\cite{LeeRama} which are nonanalytic dependences of 
zero-temperature ($T=0$) electronic 
correlation functions on frequency and/or wavenumber. Loss of phase coherence
cuts off these singularities, and transfers the nonanalyticities into 
nonanalytic dependences on the phase breaking parameter. The standard
interpretation of the weak-localization experiments is based on the
temperature dependence of various phase breaking processes and the
resulting nonanalytic temperature dependence of various observables,\cite{AAR}
and an understanding of phase breaking processes is therefore of great 
importance. Of particular interest is phase breaking due to
inelastic scattering caused by the electron-electron interaction,
since at low temperatures this dominates other temperature dependent
scattering processes, e.g., those caused by phonons.

For many years there had been consensus that at zero temperature the
phase-breaking rate due to interactions vanishes, so that in the absence of
other dephasing processes like, e.g., magnetic scattering, the electrons
maintain full interference capabilities. This consensus was based on
calculations of various phase-breaking times. An important example is the
single-particle inelastic life time $\tau_{\rm in}$, which describes the
decay of the single-particle Green function,
\be
G({\bf p},i\omega_n) = \frac{Z^{1/2}}{-i\omega_n + \xi_{\bf p} 
                       + i\,\sgn (\omega_n)/2\tau_{\rm in}}\quad.
\label{eq:1.1}
\ee
Here ${\bf p}$ is the wavenumber, $\xi_{\bf p} = \epsilon({\bf p}) - \mu$
with $\epsilon({\bf p})$ the renormalized single-particle energy and 
$\mu$ the chemical potential, $\omega_n = 2\pi T (n+1/2)$ is a fermionic 
Matsubara frequency, and $Z$ is the wavefunction renormalization. Landau 
Fermi-liquid theory predicts the single-particle inelastic scattering rate, 
$1/\tau_{\rm in}$, in clean electron fluids to vanish like 
$1/\tau_{\rm in}\propto T^2$.\cite{BaymPethick}
Schmid calculated $1/\tau_{\rm in}$ for disordered systems with diffusive
electron dynamics, and found it to be enhanced compared to the clean case,
vanishing like $T^{d/2}$ in $2<d<4$ spatial dimensions.\cite{Schmid}
The transport properties
measured in weak-localization experiments are governed by a similar
relaxation time, namely, the phase breaking time $\tau_{\phi}$ that
is related to a two-particle correlation function often referred
to as the diffuson propagator,\cite{CDKL,CKL}
\be
L({\bf p},i\Omega_n) = \frac{Z}{\Omega_n{\wt H} + D{\bf p}^2 
                       + 1/\tau_{\phi}}\quad.
\label{eq:1.2}
\ee
Here $\Omega_n = 2\pi Tn>0$ is a bosonic Matsubara frequency, ${\wt H}$ is a
frequency renormalization, and $D$ is a renormalized diffusion coefficient.
For $2<d<4$, many different calculation found the temperature dependences 
of $\tau_{\phi}$ and $\tau_{\rm in}$ to be qualitatively the same, 
$1/\tau_{\phi}\propto 1/\tau_{\rm in}\propto T^{d/2}$.\cite{tau_footnote}
In a weak-localization context, instead of the diffuson, the closely related
Cooperon propagator is often discussed, which has a mass that in time-reversal
invariant systems is also given by $\tau_{\phi}$.\cite{AAK}
An overview within a unifying theoretical framework has been given in
Ref.\ \onlinecite{ChakravartySchmid}.

This consensus was shattered in the wake of experiments on disordered
metallic wires that found that $1/\tau_{\phi}$ saturates as 
$T\rightarrow 0$.\cite{MJW,LinBird} 
The experimentalists assert to have ruled out 
other sources of dephasing, which suggests that
the electron-electron interaction can lead to residual dephasing even at
$T=0$. These experiments deal with finite-size one-dimensional or 
two-dimensional samples only, and for these systems an explanation in
terms of quantum fluctuations has been proposed.\cite{MohantyWebb}
However, more detailed theoretical work along the same lines
found such a residual contribution to $1/\tau_{\phi}$ not just in $d=1$ 
and $d=2$, but also in $d=3$,\cite{GolubevZaikin} and all of these results
are in contradiction to the earlier theoretical work.\cite{AAK} 
The quantum fluctuations that are proposed to lead to the saturation have 
been taken into account in the earlier
calculations, as Ref.\ \onlinecite{ChakravartySchmid} makes explicit, yet
the results disagree. This contradiction has recently given rise to 
an intense debate.\cite{GZS,AAV} It is important to note that this
disagreement by far transcends the technical issue of what the behavior
of a particular observable is, and how to calculate it. A finite dephasing
rate at $T=0$ in three-dimensions would mean the breakdown of a Fermi-liquid 
picture for disordered bulk Fermi systems, with far-reaching consequences.

In this paper we attack the problem from a very general theoretical angle.
Rather than calculating the phase-breaking rate explicitly, we use general
symmetry arguments and Ward identities to prove that the electron-electron
interaction cannot give rise to a nonvanishing value of $1/\tau_{\phi}$ at
zero temperature in $d>2$. This mathematical result establishes that the
diffuson propagator is always massless at zero temperature. We then show
that in the presence of time reversal invariance, the same is true for
the Cooperon propagator.

\section{Spontaneous symmetry breaking and soft modes in electron systems}
\label{sec:II}

The discussion in the literature of conflicting results obtained by 
different calculations that we summarized in the Introduction fails to 
mention an important point. Namely, the question of the
existence or otherwise of a finite electron dephasing time at zero temperature
is a question about the soft-mode spectrum of the system, since a nonzero
dephasing rate gives certain propagators a mass, see Eq.\ (\ref{eq:1.2}). 
The existence and identity
of soft modes, however, is governed by fundamental properties of the system
and can be studied rigorously, and in very general terms, without reference
to any particular explicit calculation. In fact, such an analysis should
always be performed {\em before} carrying out calculations, since
the former provides important checks for the latter. This is particularly
important if the calculations involve approximations, as it is
crucial to make sure that the approximations are `conserving' in the sense
that they do not violate any exact qualitative
properties. Of course, systematic perturbation theory, if done correctly,
is always conserving in this sense, as any exact properties will hold order
by order in perturbation theory. However, given the complexity of many
perturbative calculations, and the fact that perturbation theory by itself
can never establish that some property is actually exact, it is still valuable 
to have exact benchmarks.

There are two known mechanism for guaranteeing that certain correlation
functions are massless, viz., (1) conservations laws, and (2) spontaneously
broken symmetries. Conservation laws ensure that a conserved quantity cannot
be locally created or destroyed. An excitation that couples to a conserved
quantity can only decay by some of the quantity being transported into or
out of the
perturbed region. For long-wavelength excitations, this takes a long time.
Therefore, the correponding correlation functions will decay infinitely
slowly in the limit of zero wavenumber and zero frequency, i.e., they are
massless. An example are density fluctuations in any isolated many-particle
system, whose decay rate goes to zero with vanishing wavenumber due to the 
particle number conservation law. This is of course true both at zero and
at nonzero temperature. A spontaneously broken continuous symmetry also leads 
to soft modes, the number of which depends on the number of parameters of the
symmetry group that is broken, and on the extent of the breaking. This is
the content of Goldstone's theorem, and the resulting soft modes are usually
referred to as Goldstone modes.\cite{Forster} An example are the spin waves
in a Heisenberg antiferromagnet, which are the Goldstone modes resulting from
the spontaneously broken spin rotation symmetry.

The systematic analysis of symmetries and the resulting soft modes in
disordered itinerant electron systems was pioneered by Wegner,\cite{Wegner}
and by McKane and Stone.\cite{McKaneStone} These authors showed that, for
noninteracting electrons, the diffusive excitations that give rise to the
weak-localization effects are the Goldstone modes of a spontaneously broken
symmetry that can be formulated as a rotation in frequency space. Roughly
speaking, it is the symmetry between retarded and advanced Green functions
that is spontaneously broken whenever there is a nonzero density of states.
A nonzero temperature serves as an external field conjugate to the order
parameter, i.e. the density of states, and gives the Goldstone modes a mass.
The results of Refs.\ \onlinecite{GolubevZaikin,GZS} amount to 
the claim that either, the electron-electron interaction provides such a 
conjugate field even at $T=0$, making the Goldstone modes massive, or,
that time reversal invariance is always broken in interacting many-body
systems, making the Cooperon different from the diffuson.

Wegner's symmetry analysis has been generalized to the case of interacting
electrons in Refs.\ \onlinecite{us_fermionsI,us_fermionsII}, and was used
to construct an effective field theory for disordered interacting electrons.
Although it is implicit in this work that at zero temperature, and in $d>2$,
the electron-electron interaction cannot lead to any dephasing, the controversy
discussed above makes it worthwhile to re-analyze the symmetry arguments and
prove this statement explicitly. We will now proceed to do so.

\section{Absence of phase breaking at zero temperature}
\label{sec:III}

\subsection{Field theory for fermions}
\label{subsec:III.A}

Our starting point is a general field theory for electrons. For any
fermionic system, the partition function can be written\cite{NegeleOrland}
\begin{mathletters}
\label{eqs:3.1}
\begin{equation}
Z = \int D[{\bar\psi},\psi]\ \exp\left(S[{\bar\psi},
                       \psi] \right)\quad,
\label{eq:3.1a}
\end{equation}
where $S$ is the action in terms of the fermionic (i.e., Grassmann valued)
fields $\bar\psi$ and $\psi$. We consider an action that consists of a
free-fermion part $S_0$, a part $S_{\rm dis}$ describing the interaction
of the electrons with quenched disorder, and a part $S_{\rm int}$ describing
the electron-electron interaction,
\begin{equation}
S = S_0 + S_{\rm dis} + S_{\rm int}\quad,
\label{eq:3.1b}
\end{equation}
\end{mathletters}%
Each field $\psi$ or $\bar\psi$ carries a Matsubara frequency
index $n$ and a spin index $\sigma=\uparrow,\downarrow$. In terms of
these fields, $S_0$ reads explicitly
\be
S_0 = -\int d{\bf x}\sum_{n,\sigma}{\bar\psi}_{n\sigma}({\bf x})\,
      \left[i\omega_n + \epsilon(\nabla) + \mu\right]\,\psi_{n\sigma}({\bf x})
      \quad,
\label{eq:3.2}
\ee
with $\mu$ the chemical potential, and
$\omega_n = 2\pi T(n+1/2)$ a fermionic Matsubara frequency.
$\epsilon$ denotes the dispersion relation. For instance, for free electrons
one has $\epsilon(\nabla) = -\nabla^2/2m$, with $m$ the electron mass.
Since we will be interested in four-fermion correlation functions,
it is useful to introduce a matrix of bilinear products of the fermion 
fields,\cite{ELK}
\begin{eqnarray}
B_{nm} &=& \frac{i}{2}\left( \begin{array}{cccc}
          \m-\psi_{n\uparrow}{\bar\psi}_{m\uparrow} &
            \m -\psi_{n\uparrow}{\bar\psi}_{m\downarrow} &
                \m -\psi_{n\uparrow}\psi_{m\downarrow} &
                    \m  \ \ \psi_{n\uparrow}\psi_{m\uparrow}  \\
          \m-\psi_{n\downarrow}{\bar\psi}_{m\uparrow} &
             \m-\psi_{n\downarrow}{\bar\psi}_{m\downarrow} &
                 \m-\psi_{n\downarrow}\psi_{m\downarrow} &
                    \m  \ \ \psi_{n\downarrow}\psi_{m\uparrow}  \\
          \m\ \ {\bar\psi}_{n\downarrow}{\bar\psi}_{m\uparrow} &
            \m \ \ {\bar\psi}_{n\downarrow}{\bar\psi}_{m\downarrow} &
               \m  \ \ {\bar\psi}_{n\downarrow}\psi_{m\downarrow} &
                    \m  -{\bar\psi}_{n\downarrow}\psi_{m\uparrow} \\
          \m-{\bar\psi}_{n\uparrow}{\bar\psi}_{m\uparrow} &
             \m-{\bar\psi}_{n\uparrow}{\bar\psi}_{m\downarrow} &
                \m -{\bar\psi}_{n\uparrow}\psi_{m\downarrow} &
                     \m \ \ {\bar\psi}_{n\uparrow}\psi_{m\uparrow} \\
                    \end{array}\right)
\nonumber\\
&\cong& Q_{nm}\quad.
\label{eq:3.3}
\end{eqnarray}
where all fields are understood to be taken at position ${\bf x}$. The
matrix elements of $B$ commute with one another, and are therefore
isomorphic to classical or number-valued fields that we denote by 
$Q$.\cite{Notation_Footnote}
This isomorphism maps the adjoint operation on products of fermion fields,
which is denoted above by an overbar, on the complex conjugation of the
classical fields. We use the isomorphism to
constrain $B$ to the classical field $Q$, and exactly rewrite the partition
function\cite{us_fermionsI}
\begin{eqnarray}
Z &=& \int D[{\bar\psi},\psi]\ e^{S[{\bar\psi},\psi]}
      \int D[Q]\,\delta[Q-B]
\nonumber\\
  &=& \int D[{\bar\psi},\psi]\ e^{S[{\bar\psi},\psi]}
      \int D[Q]\,D[{\wt\Lambda}]\ e^{\Tr [{\wt\Lambda}(Q-B)]}
\nonumber\\
  &\equiv& \int D[Q]\,D[{\wt\Lambda}]\ e^{{\cal A}[Q,{\wt\Lambda}]}\quad.
\label{eq:3.4}
\end{eqnarray}
${\wt\Lambda}$ is an auxiliary bosonic matrix field that serves to enforce the
functional delta-constraint in the first line of Eq.\ (\ref{eq:3.4}),
and the last line defines the action ${\cal A}$.
The matrix elements of both $Q$ and ${\wt\Lambda}$
are spin-quaternions (i.e., elements of ${\cal Q}\times{\cal Q}$ with
${\cal Q}$ the quaternion field). From Eq.\ (\ref{eq:3.3}) we see that
expectation values of the $Q$ matrix elements yield 
local Green functions, and $Q$-$Q$
correlation functions describe four-fermion correlation functions.
The physical meaning of ${\wt\Lambda}$ 
is that its expectation value plays the role of a
self energy, see Ref.\ \onlinecite{us_fermionsI}.

It is convenient to expand the $4\times 4$ matrix in Eq.\ (\ref{eq:3.3})
in a spin-quaternion basis,
\begin{equation}
Q_{nm}({\bf x}) = \sum_{r,i=0,3} (\tau_r\otimes s_i)\,{^i_rQ_{nm}}({\bf x})
                 \quad
\label{eq:3.5}
\end{equation}
and analogously for $\wt\Lambda$. Here 
$\tau_0 = s_0 = \openone_2$ is the
$2\times 2$ unit matrix, and $\tau_j = -s_j = -i\sigma_j$, $(j=1,2,3)$,
with $\sigma_{1,2,3}$ the Pauli matrices. In this basis, $i=0$ and $i=1,2,3$
describe the spin singlet and the spin triplet, respectively. An explicit
calculation reveals that $r=0,3$ corresponds to the particle-hole channel
(i.e., products ${\bar\psi}\psi$), while $r=1,2$ describes the
particle-particle channel (i.e., products ${\bar\psi}{\bar\psi}$ or
$\psi\psi$). We will be particularly interested in the matrix elements of
${^0_0 Q}$ and ${^0_1 Q}$, for which the isomorphism expressed in 
Eq.\ (\ref{eq:3.3}) reads
\bml
\label{eqs:3.6}
\begin{eqnarray}
{^0_0 Q}_{12}({\bf x})&\cong&\frac{i}{8}\sum_{\sigma}\left[
          {\bar\psi}_{1\sigma}({\bf x})\psi_{2\sigma}({\bf x})
   + {\bar\psi}_{2\sigma}({\bf x})\psi_{1\sigma}({\bf x})\right]\quad,
\nonumber\\
\label{eq:3.6a}\\
{^0_1 Q}_{12}({\bf x})&\cong&\frac{1}{8}\,\left[
   {\bar\psi}_{1\uparrow}({\bf x}){\bar\psi}_{2\downarrow}({\bf x})
  +{\bar\psi}_{2\uparrow}({\bf x}){\bar\psi}_{1\downarrow}({\bf x})
   \right.
\nonumber\\
&& + \left. \psi_{1\uparrow}({\bf x})\psi_{2\downarrow}({\bf x})
   + \psi_{2\uparrow}({\bf x})\psi_{1\downarrow}({\bf x})\right]\quad.
\label{eq:3.6b}
\eea
\eml

From the structure of Eq.\ (\ref{eq:3.3}) one obtains the
following formal symmetry properties of the $Q$ 
matrices,\cite{us_fermionsI,us_fermionsII}
\begin{mathletters}
\label{eqs:3.7}
\begin{eqnarray}
{^0_r Q}_{nm}&=&(-)^r\,{^0_r Q}_{mn}\quad,\quad (r=0,3)\quad,
\label{eq:3.7a}\\
{^i_r Q}_{nm}&=&(-)^{r+1}\,{^i_r Q}_{mn}\ ,\ (r=0,3;\ i=1,2,3)\quad,
\label{eq:3.7b}\\
{^0_r Q}_{nm}&=&{^0_r Q}_{mn}\quad,\quad (r=1,2)\quad,
\label{eq:3.7c}\\
{^i_r Q}_{nm}&=&-{^i_r Q}_{mn}\quad,\quad (r=1,2;\ i=1,2,3)\quad,
\label{eq:3.7d}\\
{^i_r Q}_{nm}^*&=&- {^i_r Q}_{-n-1,-m-1}\quad.
\label{eq:3.7e}
\end{eqnarray}
\end{mathletters}%
Here the star in Eq.\ (\ref{eq:3.7e}) denotes complex conjugation.

By using the delta constraint in Eq.\ (\ref{eq:3.4}) to rewrite all terms 
that are quartic in the fermion field in terms of $Q$, we can achieve
an integrand that is bilinear in $\psi$ and $\bar\psi$. The Grassmannian
integral can then be performed exactly, and we obtain for the
action ${\cal A}$
\begin{mathletters}
\label{eqs:3.8}
\begin{eqnarray}
{\cal A}[Q,{\wt\Lambda}] &=& {\cal A}_{\rm dis} + {\cal A}_{\rm int}
                           + \frac{1}{2}\,\Tr\ln\left(G_0^{-1} - i{\wt\Lambda}
                                       \right)
\nonumber\\
  && + \int d{\bf x}\ \tr\left({\wt\Lambda}({\bf x})\,Q({\bf x})\right)\quad.
\label{eq:3.8a}
\end{eqnarray}
Here
\begin{equation}
G_0^{-1} = -\partial_{\tau} + \epsilon(\nabla) + \mu\quad,
\label{eq:3.8b}
\end{equation}
\end{mathletters}%
is the inverse free electron Green operator, with $\partial_{\tau}$ 
the derivative with respect to imaginary time.
$\Tr$ denotes a trace over all degrees of freedom, including the continuous
position variable, while $\tr$ is a trace over all those discrete indices that
are not explicitly shown. 

For explicit calculations, the electron-electron interaction 
${\cal A}_{\rm int}$ is usually decomposed into four pieces 
that describe the interaction
in the particle-hole and particle-particle spin-singlet and spin-triplet 
channels, respectively.\cite{AGD,us_fermionsI} 
For the purposes of the present paper, which will study exact structural
properties rather than explicit calculations, this decomposition is neither
necessary nor desirable. We therefore write the interaction part of the
action as the basic density-density interaction between electrons mediated
by a statically screened Coulomb potential. In terms of fermionic fields,
this reads
\bml
\label{eqs:3.9}
\bea
S_{\rm int} &=& -\frac{1}{2}\int d{\bf x}\,d{\bf y}\,
              v_{\rm sc}({\bf x}-{\bf y})
              \sum_{\sigma_1,\sigma_2}{\bar\psi}_{\sigma_1}({\bf x})\,
              \psi_{\sigma_1}({\bf x})\,
\nonumber\\
&&\hskip 60pt \times{\bar\psi}_{\sigma_2}({\bf y})\,
              \psi_{\sigma_2}({\bf y})\quad.
\label{eq:3.9a}
\eea
Here
\be
v_{\rm sc}({\bf x}) = \frac{e^2}{\vert{\bf x}\vert}\,
                      e^{-\kappa\vert{\bf x}\vert}\quad,
\label{eq:3.9b}
\ee
is the screened Coulomb potential, with $e$ the electron charge and
$\kappa$ the Thomas-Fermi screening wavenumber. We stress that we
use a screened interaction for convenience (and since screening
is a real physical effect) only, and that our arguments do not rely
on this. A suitable modification of the procedure in Secs.\ \ref{subsec:III.B}
and \ref{subsec:III.C} still applies if one works with a bare Coulomb
interaction, and the results are the same, see 
Ref.\ \onlinecite{us_fermionsII}.

Its Fourier transform is, in $d$ spatial 
dimensions,\cite{Coulomb_FT_footnote}
\be
v_{\rm sc}({\bf q}) = \frac{1}{N_{\rm F}}\,
                      \frac{\kappa^{(d-1)}}{[{\bf q}^2 
                                           + \kappa^2]^{(d-1)/2}}\quad,
\label{eq:3.9c}
\ee
with $N_{\rm F}$ the density of states of clean, noninteracting
electrons at the Fermi level.
In terms of the $Q$ matrices, Eq.\ (\ref{eq:3.9a}) reads
\bea
{\cal A}_{\rm int}&=& 8T\int d{\bf x}\,d{\bf y}\sum_{r=0,3}
                     \sum_{n_1,n_2,m}v_{\rm sc}({\bf x}-{\bf y})
\nonumber\\
&&\hskip 10pt\times\,{^0_r Q}_{n_1,n_1+m}({\bf x})\ 
                           {^0_r Q}_{n_2,n_2+m}({\bf y})\quad,
\label{eq:3.9d}
\eea
\eml

Finally, the interaction of the electrons with the static disorder is
given by a random potential $u({\bf x})$ that couples to the electronic
density,
\bml
\label{eqs:3.10}
\be
S_{\rm dis} = -\int d{\bf x}\ u({\bf x})\sum_{\sigma}\sum_n
              {\bar\psi}_{n,\sigma}({\bf x})\,\psi_{n,\sigma}({\bf x})\quad,
\label{eq:3.10a}
\ee
or, in terms of the $Q$,
\be
{\cal A}_{\rm dis} = 4i\int d{\bf x}\ u({\bf x})\sum_n{^0_0 Q}_{nn}({\bf x})
                     \quad.
\label{eq:3.10b}
\ee
The statistical properties of the random function $u({\bf x})$ are
governed by a distribution functional $P[u]$. 
The partition function, Eq.\ (\ref{eq:3.4}), is a functional of 
$u({\bf x})$, and any physical quantity $X$, like the free energy, or
any correlation function, needs to be averaged over the disorder according
to
\be
\{X\}_{\rm dis} = \int D[u({\bf x})]\,X[u]\,P[u]\quad.
\label{eq:3.10c}
\ee
\eml%
In Refs.\ \onlinecite{us_fermionsI,us_fermionsII} the disorder average was
performed by means of the replica trick. Here, since we are not concerned about
explicit calculations, we will formally perform the disorder average by means
of the basic definition, Eq.\ (\ref{eq:3.10c}).

The above definitions completely specify our model. We stress that the
$Q$-matrices merely provide a convenient shorthand for bilinear products
of fermion fields. It would be perfectly possible to proceed in terms of
the latter, and whether or not one introduces the matrix fields is a 
matter of taste.

\subsection{Symmetry Considerations, and a Ward identity}
\label{subsec:III.B}

A symmetry analysis of the action, Eqs.\ (\ref{eqs:3.8}) - (\ref{eqs:3.10}),
has been given in Refs.\ \onlinecite{us_fermionsI,us_fermionsII}. Here
we recapitulate those aspects of this analysis that are important for
our purposes.

Let us consider an infinitesimal transformation of the $Q$-matrices,
$Q\rightarrow TQT^{-1}$, with $T=C{\hat T}C^T$, where $C=i\tau_1\otimes s_2$,
and
\be
{\hat T}_{nm} = \left(\delta_{nn_1}\delta_{mn_2} - \delta_{nn_2}\delta_{mn_1}
                \right)\,\theta + O(\theta^2)\quad.
\label{eq:3.11}
\ee
$n_1>0$ and $n_2<0$ are fixed frequency indices that characterize the
transformation. This amounts to a rotation in frequency space by an
infinitesimal angle $\theta$. These rotations are elements of the 
symplectic symmetry group ${\rm Sp}(8N)$ over the complex numbers, 
with $2N$ the number of frequency labels, which 
governs the symmetry properties of the action.\cite{us_fermionsI}
Under such a rotation, the $Q$-matrices transform like
\bml
\label{eqs:3.12}
\be
Q_{nm} \rightarrow Q_{nm} + \delta Q_{nm}\quad,
\label{eq:3.12a}
\ee
with
\be
\delta Q_{nm} = \left(\delta_{nn_1}Q_{n_2m} + \delta_{mn_1}Q_{nn_2} -
                (1\leftrightarrow 2)\right)\,\theta + O(\theta^2)\ .
\label{eq:3.12b}
\ee
\eml
The $\wt\Lambda$ transform analogously. The symbol $(1\leftrightarrow 2)$ 
denotes the same terms as written previously,
but with the indices $1$ and $2$ interchanged.

Of the terms in the action, Eq.\ (\ref{eq:3.8a}), only ${\cal A}_{\rm int}$
and the $\Tr\ln$ term are not invariant under these rotations. To linear
order in the transformation parameter $\theta$ one finds
${\cal A} \rightarrow {\cal A} + \delta{\cal A}$, with
\bml
\label{eqs:3.13}
\be
\delta{\cal A} = \frac{\theta}{2}\,\Tr (G\,\delta i\omega) 
                 + \delta{\cal A}_{\rm int}\quad,
\label{eq:3.13a}
\ee
with $G \equiv (G_0^{-1} - i\tilde\Lambda)^{-1}$, and
\bea
(\delta i\omega)_{nm}^{\alpha\beta}&=& 
   \left(\delta_{n n_1}\delta_{m n_2} +
         \delta_{n n_2}\delta_{m n_1}\right)\,
                       i\Omega_{n_1 - n_2}\quad,
\label{eq:3.13b}\\
\delta {\cal A}_{\rm int} &=& 32 \int d{\bf x} d{\bf y}\,
                 v_{\rm sc}({\bf x-y}) \sum_{r=0,3}
               T\sum_{{n}_a {n}_b}\Bigl[
               {^{0}_{r}Q}_{{n}_a {n}_b}({\bf x})\,
\nonumber\\
&&\times {^{0}_{r}Q}_{n_2,n_2-({n}_a-{n}_b)}({\bf y})
           - (1\leftrightarrow 2) \Bigr]\,\theta\quad.
\label{eq:3.13c}
\eea
\eml%
Now introduce a source $J$ for the $Q$-fields, and consider the generating
functional
\be
Z[J] = \int D[Q]\,D[\tilde\Lambda]\ e^{{\cal A} + \int d{\bf x}\
                                 \tr\bigl( J({\bf x})\,Q({\bf x})\bigr)}\quad.
\label{eq:3.14}
\ee
By performing an infinitesimal transformation on the $Q$ and the $\wt\Lambda$,
one obtains from Eq.\ (\ref{eq:3.14})
\bml
\label{eqs:3.15}
\bea
0&=&\int D[Q]\,D[\tilde\Lambda]\ \left[\delta {\cal A} + \int d{\bf x}\
       \tr\bigl(J({\bf x})\,\delta Q({\bf x})\bigr)\right]
\nonumber\\
&&\hskip 50pt   \times e^{{\cal A} + \int d{\bf x}\
            \tr\bigl(J({\bf x})\,Q({\bf x})\bigr)}\quad.
\label{eq:3.15a}
\eea
Differentiating this identity with respect to $^0_0J_{n_4 n_3}$, $n_3>0$,
$n_4<0$, and putting $J=0$ yields
\be
0 = \Bigl\langle\delta {\cal A}\ {^0_0Q}_{n_1 n_2}({\bf x})\Bigr\rangle
     + \Bigl\langle{^0_0(\delta Q)}_{n_1 n_2}({\bf x}) \Bigr\rangle\quad,
\label{eq:3.15b}
\ee
\eml%
Here the angular brackets denote an average with respect to the action
${\cal A}$. From Eq.\ (\ref{eq:3.13a}) we see that the first term involves
an average $\langle GQ\rangle$. By using the identity\cite{us_fermionsI}
$\langle GQ\rangle = -2i\langle Q^2\rangle$, and performing the disorder
average, Eq.\ (\ref{eq:3.10c}), we finally obtain
\bml
\label{eqs:3.16}
\be
8\Omega_{n_1-n_2}D_{n_1n_2,n_3n_4} = \delta_{n_1n_3}\delta_{n_2n_4}\,N_{n_1n_2}
   - W_{n_1n_2,n_3n_4}\quad.
\label{eq:3.16a}
\ee
Here
\bea
N_{n_1n_2}&=&\Bigl\{\Bigl\langle{^{0}_{0}Q}_{n_1n_1}({\bf x})\Bigr\rangle
      - \Bigl\langle{^{0}_{0}Q}_{n_2n_2}({\bf x})\Bigr\rangle\Bigr\}_{\rm dis}
         \quad,
\nonumber\\
\label{eq:3.16b}\\
D_{n_1n_2,n_3n_4}&=&\int d{\bf y}\ \Bigl\{
                \Bigl\langle{^{0}_{0}(\Delta Q)}_{n_1n_2}({\bf y})\,
\nonumber\\
&&\hskip 40pt \times
                {^{0}_{0}(\Delta Q)}_{n_3n_4}({\bf x})\Bigr\rangle
                \Bigr\}_{\rm dis}\quad,
\label{eq:3.16c}
\eea
with $\Delta Q = Q - \langle Q\rangle$, and
\bea
W_{n_1n_2,n_3n_4}&=&-32 v_{\rm sc}({\bf k}\rightarrow 0)\,N_{n_1n_2}
   T\sum_{n_a,n_b} \delta_{n_a-n_b,n_1-n_2}\,
\nonumber\\
&&\hskip 100pt \times D_{n_an_b,n_3n_4}
\nonumber\\
&&-32\sum_{r=0,3}\int d{\bf y}\,d{\bf z}\,v_{\rm sc}({\bf y}-{\bf z})\,T
     \sum_{n_a,n_b}
\nonumber\\
&&\hskip -28pt\times \biggl[\Bigl\{\Bigl\langle
   {^{0}_{r}(\Delta Q)}_{n_1,n_2-(n_a-n_b)}({\bf -q})\
                 {^{0}_{r}(\Delta Q)}_{n_a n_b}({\bf q})
\nonumber\\
&&\qquad\hskip -28pt\times {^{0}_{0}(\Delta Q)}_{n_3n_4}({\bf x})\Bigr\rangle
   \Bigr\}_{\rm dis} - (1\leftrightarrow 2)\biggr] \quad.
\label{eq:3.16d}
\eea
\eml%
The physical meaning of the propagator $D$ is a phase space density correlation
function. The homogeneous, static density susceptibility $\chi$ can be obtained
from $D$ by the relation\cite{us_fermionsI} $\chi = 16T\sum_{n,m} D_{nn,mm}$.

The Ward identity,\cite{us_fermionsI,us_fermionsII} Eq.\ (\ref{eq:3.16a}),
relates the two-point
$Q$-correlation function $D$ to the one-point function $N$ and the
three-point function $W$. Various aspects of this identity have been
discussed in Refs.\ \onlinecite{us_fermionsI,us_fermionsII}. In the next
subsection we focus on the implications for the dephasing time.

\subsection{Absence of Dephasing}
\label{subsec:III.C}

We now use the Ward identity, Eq.\ (\ref{eq:3.16a}), to show that there cannot
be a nonzero phase relaxation rate at zero temperature in $d>2$. We will 
first concentrate on the diffuson propagator, and then show that, in time
reversal invariant systems, the same conclusions hold for the Cooperon
propagator.

\subsubsection{The diffuson propagator}
\label{subsubsec:III.C.1}

Let us consider the structure of the Ward identity, Eq.\ (\ref{eq:3.16a}).
Equation (\ref{eq:3.6a}) shows that $\langle\,{^0_0 Q}_{nn}({\bf x})\rangle$ is 
proportional to the on-site electron Green function, so for $n_1>0$, $n_2<0$, 
$\Omega_{n_1-n_2}\equiv\omega_{n_1}-\omega_{n_2}\rightarrow 0$, $N_{n_1n_2}$ 
approaches a constant
$N$ that is proportional to the density of states at the Fermi level.
$D$ and $W$ can be decomposed into two pieces each that have
different frequency structures,
\bea
D_{n_1n_2,n_3n_4}&=&\delta_{n_1,n_3}\,\delta_{n_2,n_4}\,D^{\rm (dc)}_{n_1n_2}
\nonumber\\
&&\hskip 20pt + \delta_{n_1-n_2,n_3-n_4}\,D^{\rm (c)}_{n_1n_2,n_3n_4}
                    \quad,
\label{eq:3.17}
\eea
and analogously for $W$. The superscripts `(c)' and `(dc)' refer to
`connected' and `disconnected' (at a fermionic level) contributions. 
The disconnected piece of $D$, $D^{\rm (dc)}$, is identical, apart from
a proportionality constant, with the homogeneous limit $({\bf p}\rightarrow 0)$
of the diffuson propagator $L$, 
Eq.\ (\ref{eq:1.2}), that was defined diagrammatically in
Ref.\ \onlinecite{CDKL}.
For noninteracting electrons, the connected piece of $D$ vanishes,
and we have, in the limit $\Omega_{n_1-n_2}\rightarrow 0$,
\be
D^{\rm (dc)}_{n_1n_2} \rightarrow N/8\Omega_{n_1-n_2}\quad.
\label{eq:3.18}
\ee
This expresses the undisputed fact that in the absence of an electron-electron
interaction, the diffuson is massless at zero 
temperature.\cite{finite_T_footnote}

The question is now whether the presence of $W$ can change this fact. The
only possibility is a cancellation between $N$ and $W$ in the limit of
zero frequency and zero temperature. To see that this cannot happen,
let us consider the equation for $D^{\rm (dc)}$ in this case. It
is sufficient to show that $D^{\rm (dc)}$ remains massless, since
$D^{\rm (c)}$ and $D^{\rm (dc)}$ cannot cancel one another due to their 
different frequency structures. From Eq.\ (\ref{eq:3.16a}) we obtain
\bml
\label{eqs:3.19}
\be
8\Omega_{n_1-n_2}\,D^{\rm (dc)}_{n_1n_2} = N - W^{\rm (dc)}\quad,
\label{eq:3.19a}
\ee
An evaluation of Eq.\ (\ref{eq:3.16d}) yields
\bea
W^{\rm (dc)}&=&8\lim_{\omega_{n_1}\rightarrow 0+\atop\omega_{n_2}\rightarrow 0-}
     \int d{\bf y}\ d{\bf z}\ v_{\rm sc}({\bf y}-{\bf z}) T\sum_{n_a,n_b}
\nonumber\\
&&\hskip - 20pt\times\biggl\{
   G({\bf x},{\bf z};\omega_{n_1})\biggl\langle \left({\bar\psi}_{n_a}({\bf y})
                            s_0 \psi_{n_b}({\bf y})\right)
\nonumber\\
&&\qquad\hskip -20pt\times\left(({\bar\psi}_{n_2-n_a+n_b}({\bf z}) s_0 
         \psi_{n_2}
    ({\bf x}) \right)\biggr\rangle^c - (1\leftrightarrow 2)\biggr\}_{\rm dis}
                                                                  \ .
\nonumber\\
\label{eq:3.19b}
\eea
\eml%
Here we have introduced two-component spinors, $\psi_n({\bf x}) =
(\psi_{n\uparrow}({\bf x}),\psi_{n\downarrow}({\bf x}))$, and a scalar
product in spinor space that is given by the usual matrix product.
The $\langle{\bar\psi}\psi{\bar\psi}\psi\rangle^c$ are connected
$\psi$-correlation functions, with the cumulant taken with respect to
the quantum mechanical average only, and
$G({\bf x},{\bf z};i\omega_n) = \langle {\bar\psi}_{n\sigma}({\bf x})\,
\psi_{n\sigma}({\bf z})\rangle$ is the Green function for a given
disorder configuration.

Now consider $W^{\rm (dc)}$, Eq.\ (\ref{eq:3.19b}), as a function of the
screening wavenumber $\kappa$. The behavior for small $\kappa$ is determined
by the infrared properties of the integral that defines $W^{\rm (dc)}$.
The connected four-fermion correlation
functions contain at least one screened interaction potential, and
any four-fermion function is at most diffusive, i.e. diverges at most like
an inverse wavenumber squared for small wavenumbers. Counting the number
of frequency and momentum integrals in Eq.\ (\ref{eq:3.19b}), we conclude
that the leading zero-temperature contribution to $W^{\rm (dc)}$ for 
$\kappa\rightarrow 0$ is bounded above by
\bea
W^{\rm (dc)}&<&{\rm const.} \int_0^{\Lambda} dp\ p^{d-1}\ 
               \frac{\kappa^{2(d-1)}}{[\kappa^2 + p^2]^{d-1}}
               \ \frac{1}{p^2} 
\nonumber\\
&\propto&\kappa^{d-2} + o(\kappa^{d-2})\quad,
\label{eq:3.20}
\eea
where $o(x)$ denotes terms that vanish faster than $x$ as $x\rightarrow 0$.
$\Lambda$ is an ultraviolet cutoff that is on the order of the Fermi 
wavenumber.
It follows that $W^{\rm (dc)}$ vanishes continuously as $\kappa\rightarrow 0$
for $d>2$. The density of states, in contrast, approaches a finite value,
and so $N = O(1)$ in this limit. We see that $N$ and $W^{\rm (dc)}$ are
different functions of $\kappa$, so in general they cannot cancel. This
means that the diffuson propagator remains soft in the presence of 
interactions. It can become massive at most at a special value of the
screening wavenumber, which would signalize a phase transition into a
non-Fermi liquid state. If one replaces the screened Coulomb interaction by a
different short-ranged model interaction with interaction amplitude 
$\Gamma$,\cite{CDLM} then
all of the above arguments still apply if one considers $N$ and $W^{\rm (dc)}$
as functions of $\Gamma$ instead of $\kappa$.

The realization that $W$ and $N$ are necessarily different functions of
the interaction, and therefore cannot generically cancel, is the most
important ingredient of our proof. Notice that a cancellation would require
$W^{\rm (dc)}$ to be of $O(1)$ for $\kappa\rightarrow 0$. It follows from
Eq.\ (\ref{eq:3.20}) that the only way for this to happen is to have
a stronger than diffusive infrared singularity in a four-fermion propagator.
However, the nonvanishing phase relaxation rate that might result from this
would make the diffusive Goldstone modes
massive, i.e., it would lead to a theory that is {\em less} singular
in the infrared. Any scenario that has the diffuson become massive due
to the electron-electron interaction would
therefore not even be internally consistent.

We finally note that no cancellation between $N$ and $W^{\rm (dc)}$ is 
necessary to produce dephasing at a nonzero temperature. According to the 
relation between $D^{\rm (dc)}$ and $L$ from Eq.\ (\ref{eq:1.2}) mentioned 
above, and after an analytic continuation to real frequencies $\Omega$, one has
\be
N - W^{\rm (dc)} = \frac{\rm const.}{{\tilde H} 
                     + i/\Omega\tau_{\phi}} \quad.
\label{eq:3.19'}
\ee
As long as, for $T\rightarrow 0$ and $\Omega\rightarrow 0$, 
a regime exists where $\Omega\tau_{\phi} >> 1$, Eq.\ (\ref{eq:3.19'})
allows for an expansion in powers of $1/\Omega\tau_{\phi}$, with
the leading contribution to $N - W^{\rm (dc)}$ a nonzero constant. This is
the case if $1/\tau_{\phi}$ vanishes continuously as $T\rightarrow 0$, and
in particular if $\tau_{\phi} \propto T^{d/2}$. However, in order to
have $\tau_{\phi} \rightarrow {\rm const.}$ for $T\rightarrow 0$, the leading
$O(1)$ contribution must cancel, which is generically impossible as shown
above.

\subsubsection{The Cooperon propagator}
\label{subsubsec:III.C.2}

Finally, we return to the Cooperon propagator, which in
our language is given by the correlation function
\bea
C_{n_1n_2,n_3n_4}&=&\int d{\bf y}\ \Bigl\{
                \Bigl\langle{^{0}_{1}(\Delta Q)}_{n_1n_2}({\bf y})\,
\nonumber\\
&&\hskip 40pt \times
                {^{0}_{1}(\Delta Q)}_{n_3n_4}({\bf x})\Bigr\rangle
                \Bigr\}_{\rm dis}\quad,
\label{eq:3.21}
\eea
The decomposition analogous to Eq.\ (\ref{eq:3.17}) reads in this case
\bea
C_{n_1n_2,n_3n_4}&=&\delta_{n_1,n_3}\,\delta_{n_2,n_4}\,C^{\rm (dc)}_{n_1n_2}
\nonumber\\
&&\hskip 20pt + \delta_{n_1+n_2,n_3+n_4}\,C^{\rm (c)}_{n_1n_2,n_3n_4}
                    \quad,
\label{eq:3.22}
\eea
From Eq.\ (\ref{eq:3.6b}) we see that the disconnected piece can be expressed 
in terms of Green functions as
\bml
\label{eqs:3.23}
\be
C^{\rm (dc)}_{nm} = \frac{-1}{16}\sum_{{\bf k},{\bf p}}\left\{
                        {\cal G}_{{\bf k},{\bf p}}(i\omega_n)\,
                        {\cal G}_{-{\bf k},-{\bf p}}(i\omega_m)
                                                    \right\}_{\rm dis}\quad.
\label{eq:3.23a}
\ee
Here ${\cal G}$ is the electron Green function before disorder averaging.
It is related to $G$, Eq.\ (\ref{eq:1.1}), by
$\{{\cal G}_{{\bf k},{\bf p}}(i\omega_n)\}_{\rm dis} = \delta_{{\bf k},{\bf p}}
\,G({\bf k},i\omega_n)$.
Analogously, the diffuson propagator can be written
\be
D^{\rm (dc)}_{nm} = \frac{1}{16}\sum_{{\bf k},{\bf p}}\left\{
                        {\cal G}_{{\bf k},{\bf p}}(i\omega_n)\,
                        {\cal G}_{{\bf p},{\bf k}}(i\omega_m)
                                                    \right\}_{\rm dis}\quad.
\label{eq:3.23b}
\ee
\eml%
In the presence of time reversal invariance one has the identity
\be
{\cal G}_{{\bf k},{\bf p}}(i\omega_n) = {\cal G}_{-{\bf p},-{\bf k}}(i\omega_n)
   \quad.
\label{eq:3.24}
\ee
Substituting this into Eq.\ (\ref{eq:3.23a}) leads to
\be
C^{\rm (dc)}_{nm} = -D^{\rm (dc)}_{nm}\quad,
\label{eq:3.25}
\ee
in time reversal invariant systems. This proves that the Cooperon is
massless as well.

\section{Conclusion}
\label{sec:IV}

In summary, we have considered a model of electrons in the presence of
quenched disorder. For this model, we have proved that the diffuson 
propagator in $d>2$ at $T=0$ remains massless in the presence of
an electron-electron interaction. This means that the dephasing rate due
interactions vanishes at $T=0$. In systems that are time-reversal invariant,
the Cooperon propagator is also massless. These results prove that, to
the extent that time reversal invariance holds for the systems studied
in Ref.\ \onlinecite{MJW}, the calculations and arguments given in 
Refs.\ \onlinecite{GolubevZaikin,GZS} cannot be used to explain the 
experimental observations.

\acknowledgments
We would like to thank Richard Webb for comments on the manuscript.
This work was supported by the NSF under Grant Nos. DMR-99-75259,
DMR-98-70597, DMR-01-32555, and DMR-01-32726.

\end{document}